\documentclass{elsart}

\usepackage{graphicx}

\begin{document}

\begin{frontmatter}

\title{Induced galaxy formation}

\author[label1]{V.I.~Dokuchaev}
\ead{dokuchaev@inr.npd.ac.ru}
\author[label1]{Yu.N.~Eroshenko}
\ead{erosh@inr.npd.ac.ru}
\author[label2]{S.G.~Rubin}
\ead{sergeirubin@list.ru}

\address[label1]{Institute for Nuclear Research of the Russian Academy
of Sciences, Moscow, Russia}
\address[label2]{ Moscow State Engineering Physics Institute, Moscow,
Russia, \\ Center for Cosmoparticle Physics "Cosmion", Moscow, Russia}

\begin{abstract}
We describe the model of protogalaxy formation around the cluster of
primordial black holes with a minimum extension of standard cosmological
model. Namely, it is supposed, that a mass fraction of the universe
$\sim10^{-3}$ is composed of the compact clusters of primordial (relict)
black holes produced during the phase transitions in the early universe. 
These clusters are the centers of the dark matter (DM) condensations.  As a
result the protogalaxies with a mass $2\cdot10^{8}M_{\odot}$ form at the
redshift $z=15$. These induced protogalaxies contain the central black holes
of mass $\sim10^5M_{\odot}$ and look like the dwarf spheroidal galaxies with
a central density spike.  Subsequent merging of the induced protogalaxies and
ordinary DM haloes leads to the standard scenario of the large scale
structure formation.  Black holes merging gives the nowadays supermassive
black holes and reproduces the observed correlations between their masses and
velocity dispersions in the bulges.
\end{abstract}

\end{frontmatter}

\section{Introduction}

The problem of galaxy formation with a supermassive central black hole (BH)
becomes more and more intriguing and ambiguous in view of the discovery of
distant quasars at redshifts $z>6$ in Sloan Digital Sky Survey \cite{z6}. 
The maximum observed value of the red-shift $z=6.41$ belongs to the quasar
with a luminosity corresponding to the accretion onto BH with a mass
$3\cdot10^{9}M_{\odot}$ \cite{will03}.  The early formation of BHs with a
mass $\sim10^{9}M_{\odot}$ meets a serious problem in the standard scenario
of massive BH formation due to the dissipative evolution of dense star
clusters \cite{dokrev}, supermassive stars or gaseous disks \cite{el2}. For
this reason the scenario with the massive black holes (PBHs)
\cite{zeld67,carr75,quin,Ru1} become more attractive.  These PBHs can be the
centers of baryonic \cite{Ryan} and dark matter (DM) \cite{DokEroPAZH}
condensation into the growing protogalaxies.  It seems probable that this
scenario of galaxy formation could coexist with the standard one.  It is
supposed that both the ordinary normal galaxies without a large central BH
and induced protogalaxies around PBHs are both formed in the universe. Their
subsequent multiple merging leads to the observable large scale
structure.

An effective mechanism of massive PBHs formation and their clusterization was
developed in \cite{Ru1,Ru2,KR04}.  As the basic example in this mechanism a
scalar field with the tilted Mexican hat potential is used. The properties of
the resulting PBH clusters appear to be strongly dependent on the value of an
initial phase of the scalar field. In addition they depend on the tilt of the
potential and the scale of symmetry breaking $f$ at the beginning of the
inflation.  This clusters of PBHs could provide the initial density
perturbations for induced protogalaxy formation.

\begin{figure}[t]
\begin{tabular}
[c]{cc}
\includegraphics[angle=0,width=0.48\textwidth]{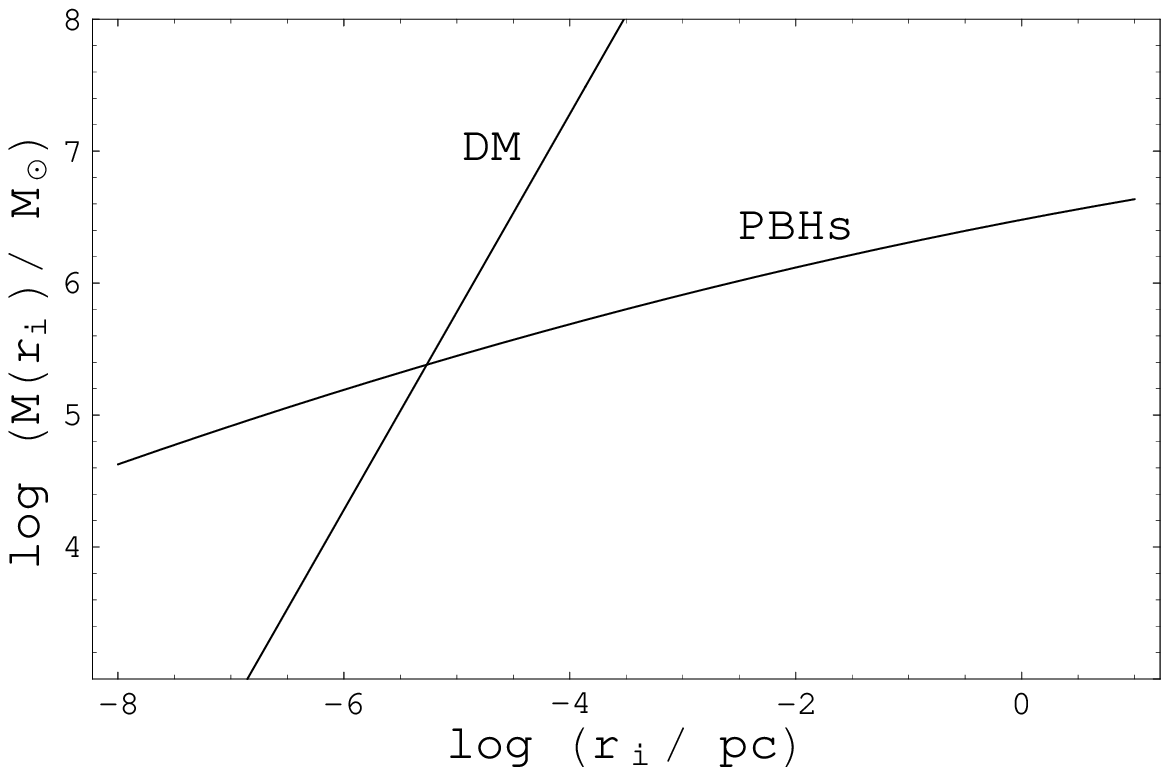}
\includegraphics[angle=0,width=0.48\textwidth]{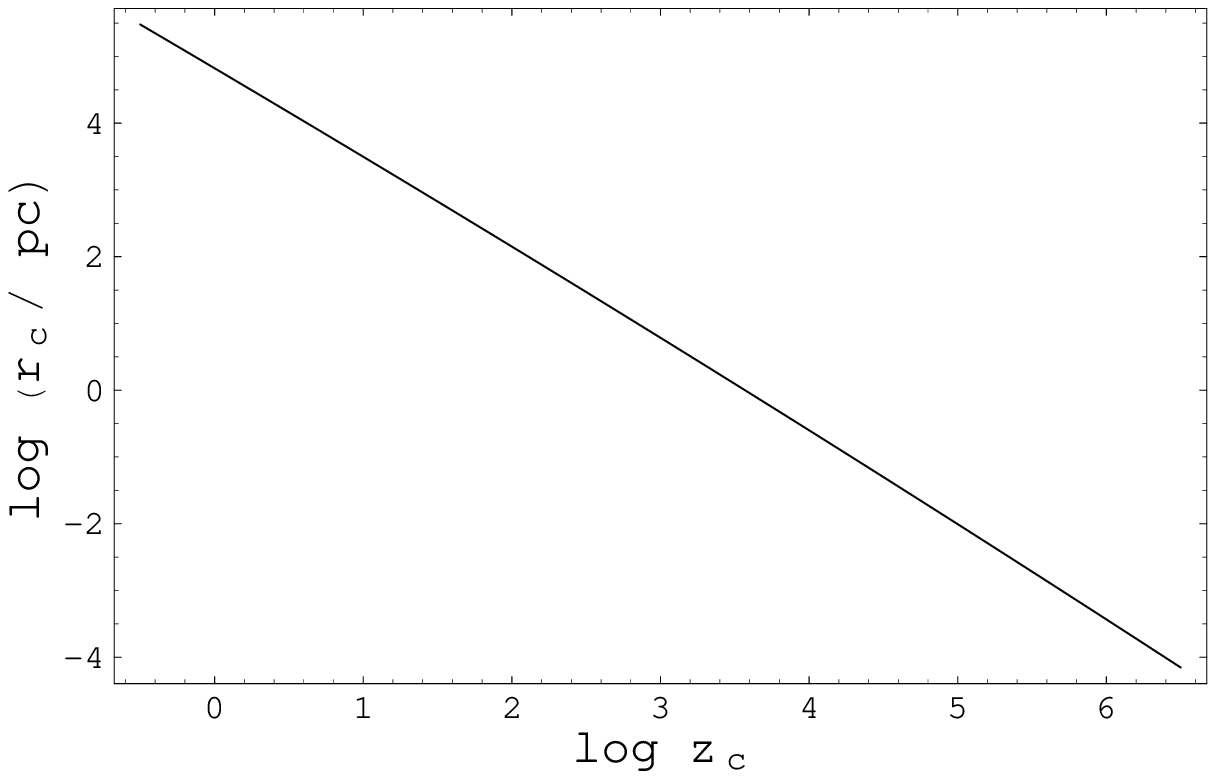}
\end{tabular}
\caption{In the left panel  are shown the initial mass profile $M_{h}(r_{i})$
of the PBHs cluster and the mass profile $M_{\rm DM}(r_{i})$ of DM. 
In the right panel the virial radius $r_{c}$ of a protogalaxy is shown as a function of
redshift $z$. }
\label{massprof}
\end{figure}

In this paper we investigate the galaxy formation around the cluster of PBHs. 
We elaborate the dynamics of DM gravitationally coupled with a PBHs cluster. 
It is shown that galaxies could be formed in this model without
any initial fluctuations in the DM. The initial mass profile $M_{h}(r_{i})$ of
the PBHs cluster is calculated following to \cite{Ru2a} and has the form
presented in the Fig.~\ref{massprof}.  For comparison the mass
$M_{\rm DM}(r_{i})$ of DM inside the same spherical shells are shown.  The
radius $r_{i}$ denotes the physical size of a sphere at the moment $t_{i}$
and the temperature $T_{i}$ when this sphere is crossing the cosmological
horizon. Note that the shells in the Fig.~\ref{massprof} are taken at
different moments $t_{i}$. Therefore the shown mass of uniformly distributed
DM doesn't follow the low $M_{\rm DM}\propto r^{3}$ as it must be for the
fixed time.  Any physical size at the temperature $T_{i}$ is smaller than
that at the current epoch at $T_{0}/T_{i}$ times, where $T_{0}=2.7$~K is the
current value of temperature.

There are several distinctive stages of BHs and galaxies formation in the
considered scenario:  (i) The formation of the closed walls of a scalar field
just after inflation and their successive collapse into the cluster of PBHs
according to \cite{Ru1,Ru2}. The formation of the most massive BH in the
center of the cluster after the horizon crossing.  (ii) The detaching of the
cluster central dense region from the cosmological expansion and its
virialization. At this stage the numerous surrounding small BHs merge
with the central BH and increase its mass.  (iii) The detaching of the cluster
outer region with the DM domination from the cosmological expansion and the
protogalaxy growth. The termination of the protogalaxy growth due to
interaction with the nearby ordinary DM fluctuations.  (iv) The cooling of
the intragalactic gas and star formation. The merging of protogalaxies
and the birth of the nowadays galaxies.

An alternative scenario could be based on the formation
large and very massive PBHs clusters.  In this case each current galaxy
contains only one PBH grown during accretion.  This possibility is 
considered in a separate paper \cite{DER1}.

\section{Gravitational dynamics of BH cluster and DM}

Let us describe the gravitational dynamics of cluster consisting of the PBHs
and DM. The results of the forthcoming calculations are applicable both for
an inner part of the cluster (composed mainly from the PBHs and collapsing at
radiation dominated stage), and for the outer regions of the cluster, where
the DM is the main dynamical component.  The outer regions of the cluster 
are detached from the cosmological expansion at the matter dominated epoch.

Consider a spherically symmetric system with a radius $r<ct$ consisting of
PBHs with a total mass $M_h$ inside the radius $r$, radiation density
$\rho_r$, DM density $\rho_{\rm DM}$ and cosmological constant density
$\rho_{\Lambda}$. The radiation density (and obviously the density of
$\Lambda$-term) is homogeneous. Therefore the fluctuations induced by PBHs
are of the type of entropy fluctuations. We use the Newton gravity because
the scales under consideration are less then the universe horizon scale. At
the same time we take into account the prescription of \cite{TolMc30} to
treat the gravitation of homogeneous relativistic components
$\rho\to\rho+3p/c^2$. The evolution of shells obeys the equation
\begin{equation}
\frac{d^2r}{dt^2}=-\frac{G(M_h+M_{\rm DM})}{r^2}-\frac{8\pi G\rho_r
r}{3}+\frac{8\pi G\rho_{\Lambda} r}{3} \label{d2rdt1}
\end{equation}
with approximate initial conditions at the moment $t_i$:  $\dot r=-Hr$,
$r(t_i)=r_i$.  During the derivation of Eq.~(\ref{d2rdt1}) it was taken
into account that $\varepsilon_r+3p_r=2\varepsilon_r$,
$\varepsilon_{\Lambda}+3p_{\Lambda}=-2\varepsilon_{\Lambda}$.  In the
parametrization $r=\xi a(t)b(t)$, the $\xi$ is a comoving length, $a(t)$ is
a scale factor of the universe normalized to the present time $t_0$ as
$a(t_0)=1$ and function $b(t)$ describes the deflection of the
cosmological expansion from the Hubble law.  Parameter $\xi$ is connected to
the mass of DM inside a considered spherical volume (excluding BHs mass) by
the relation $M_{\rm DM}=(4\pi/3)\rho_{\rm DM}(t_0)\xi^3$.  Function $a(t)$
obeys the Friedman cosmological equation, which can be rewritten as $\dot
a/a=H_0E(z)$, where redshift $z=a^{-1}-1$, $H_0$ is the current value of the
Hubble constant and function
\begin{equation}
E(z)=[\Omega_{r,0}(1+z)^4+\Omega_{m,0}(1+z)^3+
\Omega_{\Lambda,0}]^{1/2}, \label{efun}
\end{equation}
where $\Omega_{r,0}$ is a radiation density parameter,
$\Omega_{m,0}\simeq0.3$, $\Omega_{\Lambda,0}\simeq0.7$, and $h=0.7$.
By using the second Friedman equation (for $\ddot a$) one can rewrite
(\ref{d2rdt1}) in the following form
\begin{equation}
\frac{d^2b}{dz^2}+\frac{db}{dz}S(z)+
\left(\frac{1+\delta_h}{b^2}-b\right)\frac{\Omega_{m,0}(1+z)}{2E^2(z)}=0,
\label{d2bdz1}
\end{equation}
where function
\begin{equation}
S(z)=\frac{1}{E(z)}\frac{dE(z)}{dz}-\frac{1}{1+z}
\end{equation}
and the value of fluctuation $\delta_h=M_h/M_{\rm DM}$. In the limiting case
$\Omega_{\Lambda}=0$ the (\ref{d2bdz1}) is equivalent to the equation
obtained in \cite{kt}. We start tracing the evolution of cluster at a high
redshift $z_i$ when the considered shell crosses the horizon $r\sim ct$.
Initial conditions for the evolution are presented in the
Fig.~\ref{massprof}.

\begin{figure}[t]
\begin{tabular}
[c]{cc}
\includegraphics[angle=0,width=0.48\textwidth]{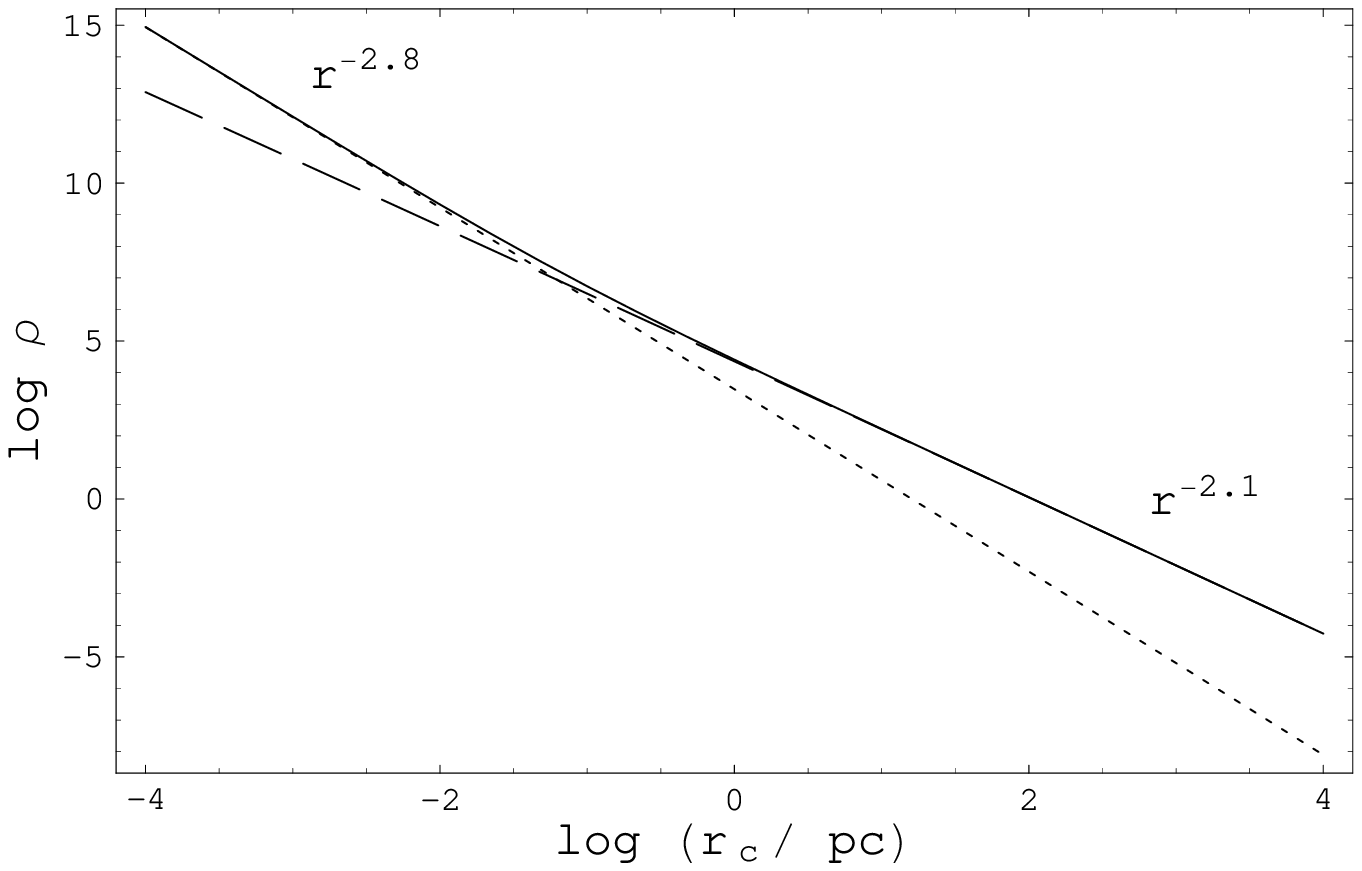}
\includegraphics[angle=0,width=0.48\textwidth]{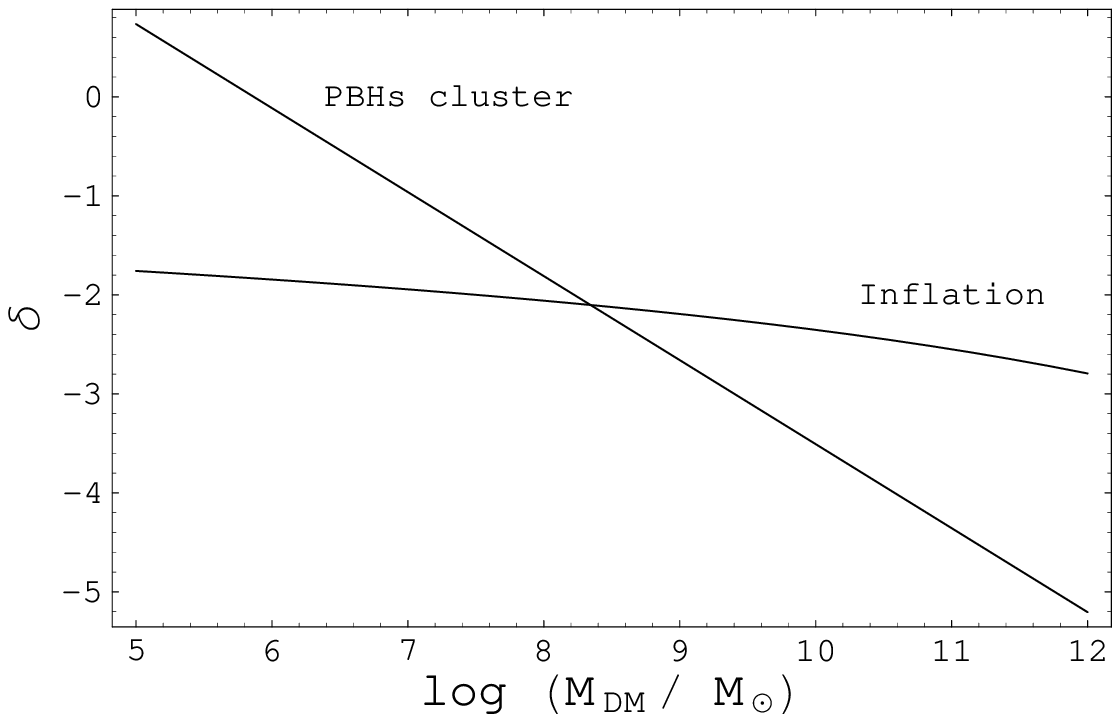} &
\end{tabular}
\caption{
In the
left panel are shown the final density profiles (\ref{dprofeq}) of a
protogalaxy ($\rho$ in the units $M_{\odot}/pc^{3}$) in dependence of the
distance from the cluster center $r_{c}$ for DM (dashed line), for BHs
(dotted line) and for their sum (solid line). The corresponding asymptotic
power laws are also labeled.
In the right panel the r.m.s. density perturbation at the moment
$t_{\rm eq}$ of matter--radiation equality produced by the cluster of PBHs
and by inflation. }
\label{solgen}
\end{figure}

The moment $t_s$ of the expansion termination $\dot{r}=0$ corresponds to
the condition $db/dz=b/(1+z)$ (according to the definition of $b$). We
accept that after the termination of expansion of specific shell, it is
virialized and contracted from maximum radius $r_{s}$ to the radius
$r_{c}=r_{s}/2$.  Therefore, average density of DM in the virialized object
$\rho$ is 8 times larger comparing with the DM density at the moment of
maximum expansion of its shell $\rho=8\rho_{m,0}(1+z_{s})^{3}b_{s}^{-3}$,
where $b_s=b(t_s)$ and an effective (virial) radius of the object
$r_{c}=(3M_{\rm DM}/4\pi\rho)^{1/3}$. Results of the numerical solution of
(\ref{d2bdz1}) are shown in the Fig.~\ref{massprof} (right panel) and
Fig.~\ref{solgen} (left panel).
Let us consider the fate of spherical shells beginning from the center of
the cluster.  It is obvious (and confirmed by numerical calculations) that
more dense inner spherical shells stop their expansion earlier than
the outer ones.  As it was discussed earlier the BH with a mass
$M_{c}=2.7\cdot10^{4}M_{\odot}$ forms in the center of the cluster at the
moment $t_{i}$. The shells near the center are very dense and are detached
from the cosmological expansion at the radiation dominated epoch.

The process similar to the 'secondary accretion' could take place for the
early formed PBHs.  As a result the PBHs would be 'enveloped' by the DM
halo.  We will call these DM haloes the induced protogalaxies.  The density
profile does not follow the secondary accretion law $\rho\propto r^{-9/4}$
because the central mass is not compact.  After virialization the distribution
of DM is
\begin{equation}
\rho_{\rm DM}(r)=\frac{1}{4\pi r^{2}}\frac{dM_{\rm DM}(r)}{dr},
\label{dprofeq}
\end{equation}
where function $M_{\rm DM}(r)$ is determined by the solution of
(\ref{d2bdz1}).
By analogy with the DM case one can obtain the profile of the BHs density and
of the total density. The results are shown in the Fig.~\ref{solgen} (left
panel), where density is expressed in units $M_{\odot}/pc^{3}$ and radial
distance is measured in parsecs.

The total mass of protogalaxy is growing with time because more and more
distant shells are detached from the cosmological expansion and are
virialized around the central most massive BH.  The growing of protogalaxy is
terminated at the epoch of nonlinear growth of the ambient density
fluctuations with the same mass $M$ as a considered system of BHs and DM. 
These fluctuations are originated in a standard way from the inflationary 
cosmological perturbations with the spectrum $P(k)$ \cite{bardeen,mo02}.  The
laws of growth for both types of fluctuations during the matter dominated
epoch are very similar. The condition for the termination of a typical
induced protogalaxy growth is
\begin{equation}
\delta_{\rm eq}^{\rm DM}(M_{\rm DM})=\delta_{\rm eq}^{h}(M_{\rm DM}).
\label{Deltaeq}
\end{equation}
The r.h.s of this equation is an amplitude of fluctuation
caused by the BH cluster.  Respectively the l.h.s. denotes the Gaussian
fluctuations.   
The value of fluctuations produced by the cluster
$\delta_{\rm eq}^{h}(M_{\rm DM})=2.5\delta_{i}(M_{\rm DM})$ is shown in
Fig.~\ref{solgen} (right panel) together with the Harrison-Zeldovich spectrum.  The factor
2.5 corresponds to the entropy perturbation growth to the moment $t_{\rm eq}$
according to the Meszaros solution.

Numerical solution of (\ref{d2bdz1}) and (\ref{Deltaeq}) gives the termination
of the protogalaxy growth at the redshift $z=15$ with a final mass of induced
protogalaxy $M_{\rm DM}=2.2\cdot10^{8}M_{\odot}$.  So the whole range of
masses and radii shown at Figures is not realized.  After this moment the
structure formation proceeds by the standard scenario:  small (ordinary and
induced) protogalaxies are assembled into the larger ones and then into the
clusters and superclusters.  Being formed, the induced protogalaxies looks
like dwarf spheroidal galaxy with a central DM spike shown in the
Fig.~\ref{solgen} and a central BH.  Some of these induced protogalaxies could
escape merging and survive till now in the form of rare dwarf galaxies.

In previous calculations we demonstrate that massive induced protogalaxy
with the mass $2.2\cdot10^{8}M_{\odot}$ is formed around the PBHs cluster at
$z=15$.  There are a lot of smaller neighboring protogalaxies, both ordinary
and induced ones, in its vicinity.  The coalescences of these protogalaxies
results in the formation of nowadays galaxies. The induced protogalaxies are
massive enough to be settled down to the galactic center during the Hubble
time under influence of the dynamical friction. The fate of BHs inside the
central few parsecs of the forming host galaxy is rather uncertain. We will
suppose the multiple BHs merging into a single BH during the Hubble time. 
Note that the dynamical friction must supply a very effective mechanism of
merging at a final stage because the density of induced protogalaxies
$\rho\propto r^{-2.8}$ strongly growing toward the center up to the small
distance $r_{h}$ from PBH where $M_{\rm DM}(r_{h})=M_{h}$
according (\ref{dprofeq}).  For example, according to our calculations $M_{\rm
DM}=7\cdot10^{4}M_{\odot}$ is contained within the radius
$r_{h}\sim8\cdot10^{-5}$~pc providing the density 
$\rho\sim2\cdot10^{15}M_{\odot}$~pc$^{-3}$.  Using the Chandrasekhar time
(see e.~g.  \cite{Sasl}) for the dynamical friction of BH with the mass
$M_{\rm BH} =7\cdot10^4 M_{\odot}$ we obtain an estimation for the
characteristic time for BH spiralling down from the radius $r_h$,
$t_{f}\sim36\mbox{~days}$.  The BHs of a larger mass would be spiralling down
even faster. As a result the late phase of BHs merging lasts very fast and so
the probability of simultaneous existence in the galactic nucleus of three or
more BHs is rather low. Thus as rule the galactic center contains only a
single supermassive BH.  On the contrary, a galaxy as a whole could contain a
substantial amount of massive PBHs which were nucleated initially far from
the galactic center \cite{Ru1,Ru2,KR04}.

Finally let us calculate the growth of the central BH under the process of the
two-body relaxation of surrounding smaller BHs in the inner region of the
induced protogalaxy. The BHs cluster is composed of the BHs of different
masses.  So the important factor of relaxation evolution is the mass
segregation: the concentration of more massive BHs at center of the cluster. 
This makes the comprehensive analytical treatment very complicated.  We use
an approximate approach by considering BHs of different masses as independent
homological subsystems evolving in a total gravitational field.  This
approach is widely used in a study of evolution of the multicomponent star
clusters.  The time of two-body relaxation is \cite{Spit}
\begin{equation}
 t_{\rm rel}\simeq\frac{1}{4\pi}\frac{v^{3}}{G^{2}m^{2}n\ln(0.4N)},
 \label{trel}
 \end{equation}
where $v\sim(GM_{\rm tot}/r_{c})^{1/2}$ is a virial velocity, $M_{\rm tot}
=M_{\rm DM}+\sum M_{\rm BH}N_{\rm BH}$, $N$ is a total number of BHs and $n$
is a number density of BHs inside the shell respectively.  In the evaporating
model \cite{Spit} the lifetime of the cluster $t_{e}\simeq40t_{\rm rel}$. 
The gravitational collapse of the remaining cluster begins when a central
part of the cluster reached the relativistic state.  A mean value of $t_{e}$
is calculated and expressed through a redshift $z$ from the relation
$t(z)=t_{e}$.  The corresponding relaxed shells are collapsing and merging
with a central BH (having initial mass $2.7\cdot10^{4}M_{\odot}$). It appears
that collapse of an additional shells gives rather small contribution the 
mass of a central BH.  Indeed, the mass of a central BH to the moment $z=15$
of  the growth termination of induced protogalaxies is $M_{\rm
BH}=6.9\cdot10^{4}M_{\odot}$. At the moment $z\simeq1.7$ (when galaxies with
the mass $M_{\rm DM}=10^{12}M_{\odot}$ are formed) the BH mass is $M_{\rm
BH}=7.2\cdot10^{4}M_{\odot}$.  If the induced galaxy is survived up to
nowadays ($z=0$), it has a central BH with the mass $M_{\rm
BH}=7.3\cdot10^{4}M_{\odot}$. So we expect that the main contributions to the
mass growing come from the accretion of surrounding media and the merging of
the central BHs of the coalesced induced protogalaxies.

\section{Discussion}

In this paper we describe the new model of galaxy formation initiated by the 
cluster of primordial black holes. Clusters of different total mass may be
formed through the mechanism of massive primordial BHs formation
\cite{KR04,Ru2a}. Respectively the described induced protogalaxies would have
the  different masses.  For numerical calculations we choose parameters of
PBHs clusters corresponding to the formation of small protogalaxies.  This
model predicts the very early epoch of galaxy and quasar formation. The other
prediction is the existence of numerous BHs of intermediate mass beyond the
dynamical centers of galaxies and in the intergalactic medium.  One of such
BHs may be observed by the Chandra telescope in the galaxy M82 \cite{Kaar00}.

More definitely, the considered model predicts the formation of induced
protogalaxy at $z=15$ with the following parameters: the halo DM mass $M_{\rm
DM}=2.2\cdot10^{8}M_{\odot}$, the virial radius $2.1$~kpc, the central BH mass
$M_{\rm BH}=7.2\cdot10^{4}M_{\odot}$ and the total BHs mass in
the protogalaxy $M_{\rm BH}=7.1\cdot10^{5}M_{\odot}$.  We suppose that large
galaxies observed now are originated through the hierarchical clustering 
of these dwarf protogalaxies.  The process of multiple merging occurs in a
stochastic manner and leads to the correlations between the central
supermassive BHs masses and galactic bulge velocity dispersions
\cite{DokEroPAZH}.  

It is worth to estimate the probability to find a nowadays galaxy without a
supermassive BH in the framework of the considered model.  The induced
protogalaxies and the most common small protogalaxies produced by the
standard inflation scenario have masses around $M_{\rm DM}=10^{8}M_{\odot}$.
Meanwhile the nowadays galaxies have masses of the order of $M_{\rm
DM}=10^{12}M_{\odot}$. Each collision of an induced protogalaxy
with an protogalaxy originated from the standard inflation perturbations
produces protogalaxy of the next generation with a massive
BH in its center.  So $10^{4}$ collisions have been happened up to now.
Let us suppose that the amount of induces protogalaxies is $\sim0.1$\%
comparing with the most common ones.  The corresponding probability to find
the galaxy without a supermassive BH is much less than $0.999^{10000}\simeq
4.5\cdot10^{-5}$.  Hence even a very small fraction of induced protogalaxies
is able to explain the observable abundance of active galactic nuclei and
quasars.

V.D and Yu.E were supported in part by the Russian Foundation for Basic
Research (RFBR) grants 02-02-16762, 03-02-16436 and 04-02-16757 and the
Russian Ministry of Science and Technology grants 1782.2003.2 and
2063.2003.2.  S.R  was supported in part by the State Contract 
40.022.1.1.1106,  by the RFBR grant 02-02-17490 and The University of Russia
grant UR.02.01.008.

\end{document}